\documentclass{article} 
\pdfoutput=1
\usepackage{nips15submit_e,times}
\usepackage{hyperref}
\usepackage{url}

\usepackage{amsmath}
\usepackage{epsfig}
\usepackage{xspace}
\usepackage{multirow}
\usepackage{amsfonts}
\usepackage{graphicx}
\usepackage{algorithm}
\usepackage{algorithmic}
\usepackage{subfigure}
\usepackage{color}
\usepackage{url} 
\usepackage{rotating} 
\usepackage{tabularx} 
\usepackage{soul,color}
\usepackage{cite}
\usepackage{pdflscape}
\usepackage{wrapfig}
\usepackage{enumitem}
\usepackage{amssymb}

\title{MaxOutProbe: An Algorithm for Increasing the Size of  Partially Observed Networks}

\author{
Sucheta Soundarajan \\
Syracuse University\\
\texttt{susounda@syr.edu} \\
\And
Tina Eliassi-Rad \\
Rutgers University \\
\texttt{tina@eliassi.org} \\
\AND
Brian Gallagher \\
Lawrence Livermore National Laboratory \\
\texttt{bgallagher@llnl.gov} \\
\And
Ali Pinar \\
Sandia National Laboratories\\
\texttt{apinar@sandia.gov} \\
}

\nipsfinalcopy 

\begin{document}

\maketitle

\begin{abstract}
Networked representations of real-world phenomena are often partially observed, which lead to incomplete networks.  Analysis of such incomplete networks can lead to skewed results.  We examine the following problem: given an incomplete network, which $b$ nodes should be  probed to bring the largest number of new nodes into the observed network?  Many graph-mining tasks require having observed a considerable amount of the network.  Examples include community discovery, belief propagation, influence maximization, etc.  For instance, consider someone who has observed a portion (say 1\%) of the Twitter retweet network via random tweet sampling.  She wants to estimate the size of the largest connected component of the fully observed retweet network.  To improve her estimate, how should she use her limited budget to reduce the incompleteness of the network?  In this work, we propose a novel algorithm, called {\sc MaxOutProbe}, which uses a budget $b$ (on nodes probed) to increase the size of the observed network in terms of the number of nodes.  Our experiments, across a range of datasets and conditions, demonstrate the advantages of {\sc MaxOutProbe} over existing methods.
\end{abstract}

\section{Introduction}
\label{sec:introduction}
Suppose that one has observed an incomplete portion $\hat{G}$ of some larger complete network $G$.\footnote{Throughout this paper, when we use the term \textit{complete}, we are referring to the completeness of the data (i.e., no information is missing), rather than to a clique structure.}   To learn more about the structure of $G$, one can probe nodes from $\hat{G}$, where each such probe reveals all the neighbors of the selected node in $G$.  The question we address here is: which nodes in $\hat{G}$ should be probed to observe as many nodes as possible in $G$?  This question is related to previous works on graph sampling and crawling.  However, unlike much of the work in graph sampling, we are not attempting to generate a sample from scratch.  Instead, we are studying how one can enhance or improve an existing incomplete network observation, without control over how it was generated or observed. Our work is motivated by problems where one only has a partial observation of the complete network; but needs the most accurate and informative picture of the complete network.  For example, suppose a network administrator who has partially observed a computer network through traceroutes.  Which parts of the observed network should she more closely examine to get the best (i.e., most complete) view of the entire network?  With a limited probing budget, how should this further exploration be done?  Alternatively,  suppose that one has obtained a sample of the Twitter network from another researcher.  The sample was (most likely) collected for some other purpose, and so may not contain the most useful structural information for one's purposes.  How should one best supplement/enrich this sampled data?

We present a novel algorithm, called {\sc MaxOutProbe},  whose goal is to select $b$ nodes in $\hat{G}$ for probing, such that the greatest number of new nodes are brought into $\hat{G}$. {\sc MaxOutProbe} achieves this goal by ranking each node $u$ in $\hat{G}$ based on its estimate of have many neighbors $u$ has outside of $\hat{G}$.  It then selects the top $b$ nodes in this ranking.  In particular, {\sc MaxOutProbe} consists of two steps: (1) \textit{Estimation} and (2) \textit{Selection}. First, during  \textit{Estimation}, for each node $u$ in $\hat{G}$, {\sc MaxOutProbe} estimates $u$'s true degree in $G$; it also estimates $G$'s average clustering coefficient.  Second, during  \textit{Selection}, {\sc MaxOutProbe} uses these two quantities to estimate the number of nodes outside $\hat{G}$ to which $u$ is adjacent.  The $b$ nodes that are estimated to have the most neighbors outside the incomplete network are selected for probing. We evaluate {\sc MaxOutProbe} on six network datasets, using incomplete networks observed by four popular sampling method, and show that {\sc MaxOutProbe} selects nodes for probing that are more valuable than those selected by comparison methods.

The \textbf{contributions} of our paper are as follows: 
\begin{itemize}
\item We present {\sc MaxOutProbe}, a two-step algorithm that first estimates graph statistics, and then selects probes in accordance with those statistics.  Unlike existing methods on estimating network characteristics, {\sc MaxOutProbe} does not assume knowledge of how the incomplete network was observed, and estimates the relevant statistics without such background knowledge.
\item Our experiments demonstrate that with respect to the task of bringing as many nodes as possible into the incomplete network, the graphs obtained by probing nodes according to {\sc MaxOutProbe} are substantially better than those by  competing methods.
\end{itemize}

\section{Problem Definition}
\label{sec:problem}
We are given an incomplete, partially observed graph $\hat{G}$ that is a part of a larger, fully observed graph $G$. We wish to gain broader observability of $G$;  that is, we wish to observe as many nodes as possible in $G$.  To aid in this task, we are allowed to probe a specified number $b$ additional nodes in $\hat{G}$ and gain more information about the probed nodes.  Thus, the problem at hand is to select the $b$ nodes which maximize the number of nodes observed in $G$ using only the structure provided in $\hat{G}$. Let $\hat{G'}$ represent the augmented graph$-$i.e., $\hat{G}$ with the information obtained from the probes.   

In this work, \emph{probing} a node $u$ is defined as learning all of $u$'s neighbors (e.g., querying Facebook for a list of all friends of a user or learning all e-mail contacts of an individual).  Moreover, there is no `master list' of nodes in $G$ that allows an algorithm to probe nodes it has not yet seen.  Thus, only nodes that already exist in $\hat{G}$ can be probed. 

Figure~\ref{fig:ProbingOverview} illustrates the probing process.   Red nodes in $\hat{G}$ have already been fully explored.  Yellow nodes are present in $\hat{G}$, but have not yet been fully explored: these are the candidates for probing.  Green nodes are not yet present in $\hat{G}$, and we have no knowledge of them.  By probing yellow nodes, we learn about the existence of the green node.  We refer to the red nodes as `fully explored', and the yellow nodes as `unexplored'.

\section{Proposed Method: MaxOutProbe}
\label{sec:method}

\begin{figure}
\centering
\includegraphics[scale=0.45]{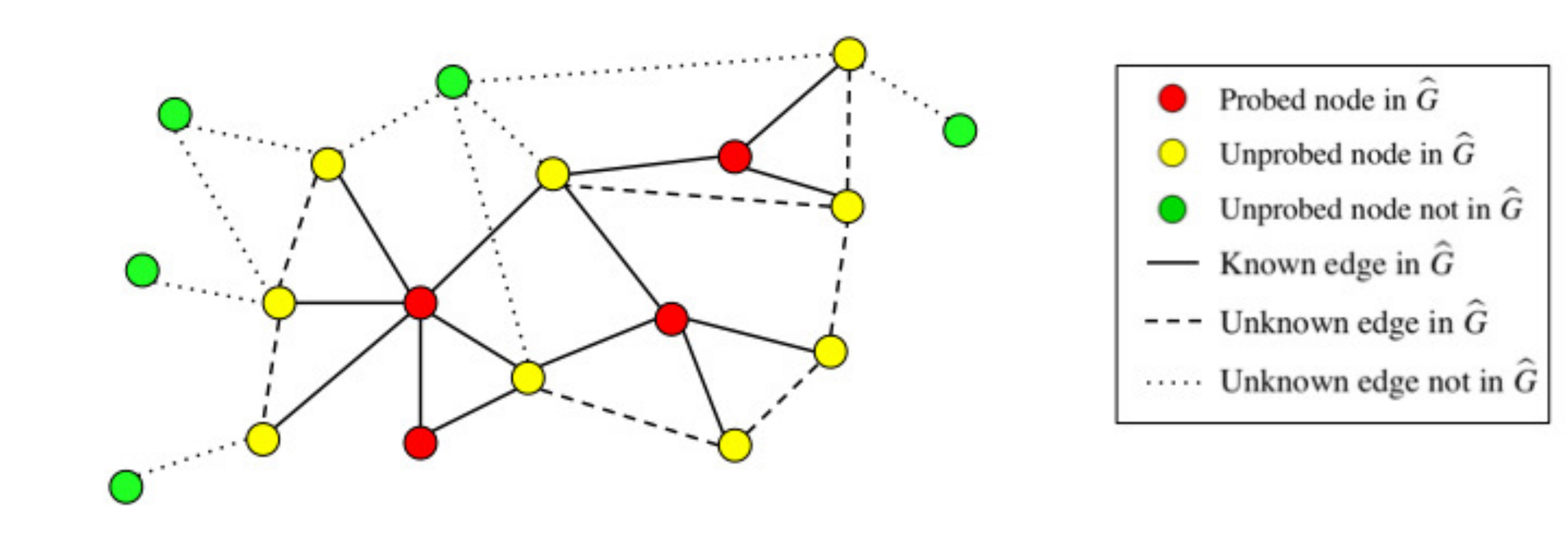}
\caption{\textnormal{Overview of probing.  Red nodes are already fully explored.  The problem is to select yellow nodes (unprobed but in $\hat{G}$) that are adjacent to many green nodes (unprobed and not in $\hat{G}$).}}
\label{fig:ProbingOverview}
\end{figure}

Given a budget of $b$ nodes to probe, {\sc MaxOutProbe} selects which $b$ unprobed nodes in $\hat{G}$ (i.e., yellow nodes in Figure~\ref{fig:ProbingOverview}) are adjacent to many nodes outside $\hat{G}$ (i.e., green nodes in Figure~\ref{fig:ProbingOverview}).  {\sc MaxOutProbe} does not make assumptions about how $\hat{G}$ was observed.

{\sc MaxOutProbe} has two steps: Estimation and Selection.  During the Estimation Step, a small amount of budget is used to probe nodes from $\hat{G}$.  The resulting information is used to estimate necessary statistics of $G$.  During the Selection Step, the statistics estimated during the Estimation Step are used to score each node in $\hat{G}$ with respect to how many neighbors outside of $\hat{G}$ that node may have. The highest scoring $b$ nodes are then selected for probing.

Let a \textit{candidate node} be a node in $\hat{G}$ that was not fully observed during the process of observing $\hat{G}$ (i.e., such a node is a candidate for probing; a yellow node in Figure~\ref{fig:ProbingOverview}).  
 For each candidate node $u$ in $\hat{G}$, {\sc MaxOutProbe} estimates $d^{out}_u$, the number of $u$'s neighbors that lie outside of $\hat{G}$, as follows:
\begin{equation}
\label{eq:in_neighbors}
d^{out}_u = d_u - d^{in}_u = d_u - d^{known}_u - d^{unknown}_u
\end{equation}

where:

\begin{itemize}[nolistsep]
\item $d_u$ is $u$'s true degree in $G$.  This quantity is unknown and must be estimated.  
\item $d^{in}_u$ is the number of nodes in $\hat{G}$ that $u$ is adjacent to in $G$.  This includes:
\begin{itemize}
\item $d^{known}_u$, the number of nodes in $\hat{G}$ that we already know to be adjacent to $u$.  This quantity can be directly calculated.
\item $d^{unknown}_u$, the number of nodes in $\hat{G}$ that $u$ is connected to in $G$, but not in $\hat{G}$ (i.e., the connections to $u$ that have not been observed).  This quantity must be estimated.  
\end{itemize}
\end{itemize}

Once $d^{out}_u$ has been calculated, {\sc MaxOutProbe} selects the $b$ nodes with the highest such values (where $b$ is the probing budget).  

We next describe how {\sc MaxOutProbe} estimates each node $u$'s true degree $d_u$ as well as the graph's average clustering coefficient $C$, which is used to estimate $d^{unknown}$ for each node.  Section~\ref{sec:selection} describes how {\sc MaxOutProbe} puts these various pieces together to obtain estimates for $d^{out}_u$.

\subsection{{\sc MaxOutProbe}'s Estimation Step}
{\sc MaxOutProbe} estimates (a) each candidate node $u$'s true degree $d_u$ and (b) the graph's clustering coefficient $C$.  Note that although there is a large body of work on estimating graph statistics, those works typically assume knowledge about how the graph was observed or generated (see Section~\ref{sec:related}).  In contrast, {\sc MaxOutProbe} makes no such assumptions. Instead to estimate the necessary statistics of $G$, {\sc MaxOutProbe} performs a series of initial probes.

\noindent\textbf{Estimating Node Degrees.}
 We empirically observe that for incomplete networks, there exists some scale factor $f$ such that multiplying $u$'s sample degree by $\frac{1}{f}$ gives a good approximation of $d_u$.  Thus, if {\sc MaxOutProbe} estimates $f$, it can estimate the true degree $d_u$ of every node in the incomplete network by using that node's observed degree.  To perform this estimate, given a total probing budget of $b$, {\sc MaxOutProbe} begins by identifying the $b$ candidate nodes from $\hat{G}$ with the highest degrees in $\hat{G}$.  A small number of these nodes are randomly selected for probing,\footnote{In our experiments, we select 100 such nodes, but this value can vary depending on the desired accuracy and total probing budget available.} and their true degrees in $G$ are learned.  By taking the average of the ratios of each selected node's true degree to its degree in $\hat{G}$, {\sc MaxOutProbe} estimates $f$.

\noindent\textbf{Estimating Average Clustering Coefficient.}
 Recall that the clustering coefficient of a node is the fraction of its neighbors that are connected, divided by the maximum possible such value.  For a candidate node $u$, {\sc MaxOutProbe} needs to know how many nodes that are presently two steps away from $u$ in $\hat{G}$ are likely to be connected to $u$ in $G$.   Thus, rather than estimating the global average clustering coefficient or the average clustering coefficient for a candidate node, {\sc MaxOutProbe} must estimate the average clustering coefficient for nodes that are adjacent to the candidate nodes.  It is the clustering coefficient of these neighboring nodes that governs the number $d^{unknown}_u$.  To make this estimation, {\sc MaxOutProbe} uses the nodes selected for probing during the estimation of $f$, as described above. For each of these selected nodes $u$, {\sc MaxOutProbe} considers each node $v$ that was two steps away in $\hat{G}$.  {\sc MaxOutProbe} then observes whether, after probing $u$, an edge connected $u$ to $v$.  By averaging these results over all selected nodes $u$, {\sc MaxOutProbe} estimates the average clustering coefficient.

\noindent\textbf{Remarks.} In cases when we know that the incomplete network was observed via random node sampling or random edge sampling, and know what fraction of the complete graph has been observed, it is possible to generate unbiased estimates for the degree of a candidate node and the clustering coefficient of the graph.  The Appendix contains the proofs for these claims.

\subsection{{\sc MaxOutProbe}'s Selection Step}
\label{sec:selection}
{\sc MaxOutProbe} uses the estimates described above to select the $b$ candidate nodes that have the most unobserved neighbors. As before, $u$ is a (candidate) node in $\hat{G}$ that was not fully explored during the observation process.  Let $f_N$ and $f_E$, respectively, denote the fraction of nodes and edges in $G$ that are present in $\hat{G}$. To estimate $d^{unknown}_u$, {\sc MaxOutProbe} uses the knowledge that real-world social and information networks tend to exhibit high clustering (i.e., have lots of triangles).  More precisely, let $w$ be an unexplored node that is two hops away from a candidate node $u$, such that $w$ and $u$ are not connected in $\hat{G}$ (but they may be connected in $G$).  $w$ forms an open wedge with $u$, because $w$ and $u$ share at least one neighbor.  Let $W_u$ be the number of such nodes $w$;  these are the nodes in $\hat{G}$ to which $u$ is most likely connected.  In the Estimation Step, {\sc MaxOutProbe} estimated the average clustering coefficient $C$ of the network, which defines the fraction of wedges that are triangles (that is, the ratio of 3 times the number of triangles in the network to the number of length-2 paths in the network).  Given this value, {\sc MaxOutProbe} estimates that $u$ is connected to $C \times W_u$ nodes in $\hat{G}$, in addition to its known neighbors in $\hat{G}$.\footnote{Although $u$'s individual clustering coefficient would be more valuable here than the global clustering coefficient $C$, {\sc MaxOutProbe} has incomplete information about $u$, and thus uses $C$ as an approximation for the per-node clustering coefficients.}  

Putting this all together, {\sc MaxOutProbe} obtains the estimate for the number of neighbors that $u$ has outside of $\hat{G}$:
\begin{equation}
\label{eq:in_neighbors}
d^{out}_u = d_u - d^{known}_u - d^{unknown}_u 
= d_u - d^{known}_u - (C \times W_u)
\end{equation}

$d^{known}_u$ and $W_u$ can be calculated exactly from $\hat{G}$.  The challenge thus lies in estimating $d_u$ and $C$, which was described in the Estimation Step above.  {\sc MaxOutProbe} computes $d^{out}_u$ for all candidate nodes in $\hat{G}$ and selects the $b$ candidate nodes with the highest $d^{out}$ values.
 
\section{Experiments}
\label{sec:exp}
Our experiments demonstrate that {\sc MaxOutProbe} outperforms a variety of other probing methods with respect to the task of maximizing the number of nodes brought into the observed network, across incomplete networks observed by different popular sampling approaches.

\noindent\textbf{Datasets.}
\label{sec:datasets}
 Table~\ref{table:datasets} describes the six real-world networks used in our experiments.  Four are communications networks, and the other two are co-occurrence networks.  The first co-occurrence network is an Amazon co-purchasing network with the nodes denoting books; and the second is a Youtube co-watching network with the nodes representing videos.

\begin{table}[t]
\centering
\begin{small}
\begin{tabular}{ c | c | c  c  c  c }
 
Type & Network $G$ & \# of Nodes & \# of Edges & $G$'s Clust.~Coeff. & \# of Conn.~Comp. \\ \hline
Communications & Enron Emails		& 84K		& 326K	&  0.08 		& 950 \\
Communications & Yahoo! IM		& 100K	& 595K	&  0.08		& 360 \\
Communications & Twitter Replies		& 261K	& 309K	&  0.002	& 11,315 \\
Communications & Twitter Retweets	& 40K		& 46K		&  0.03	 & 3,896 \\
Co-occurrences & Amazon Books	& 270K	& 741K	&  0.21		& 3840 \\
Co-occurrences & Youtube Videos	& 167K	& 1M	&  0.007		& 1 \\
\end{tabular}
\end{small}
\caption{\textnormal{Statistics for the networks considered in our experiments. Clust.~Coeff.~is short for clustering coefficient. Conn.~Comp.~is short for connected components.}}
\label{table:datasets}
\end{table}

\noindent\textbf{Sampling Methods.}
\label{sec:samplingmethods}
 We observe incomplete networks using sampling methods, each corresponding to real application scenarios.  For each sampling method, we select 10\% of the edges from the original network.  We consider the following sampling methods:
\begin{itemize}
\item \textbf{RandNode}: Nodes are sampled at random. The sample contains those nodes and all of their neighbors.  This scenario corresponds to randomly selecting online accounts and observing all of their activities (such as all friendships or communications).  This sampling method assumes that a list of observed nodes is available beforehand.  Then, one node at a time is sampled until the sample reaches 10\% of the total number of edges in the original network.  
 
\item \textbf{RandEdge}: Random edges are selected uniformly at random.  Since edges are sampled, there is no assumption that a list of observed nodes is available beforehand. This is the scenario where one randomly observes communications (such as instant messages or emails). 
 
\item  \textbf{Random Walk (RW)} and \textbf{Random Walk w/ Jump (RWJ)}:  These sampling methods begin at a random node and repeatedly transition to a random neighbor.  In RWJ, at each step there is a chance (we use 15\%) of jumping to a random node.  A single edge at a time is observed.  Thus, no assumption is made that a list of observed nodes is available beforehand.  These are common sampling methods for large networks~\cite{Leskovec2006Sampling}.
\end{itemize}  

\noindent\textbf{Other Popular Probing Methods.}
\label{sec:baselines}
 Table~\ref{table:strats} describes a variety of probing methods, each with a simple scoring function for selecting nodes.   These methods each rank all of the nodes in the incomplete graph $\hat{G}$ (except those that have already been fully explored).  Some of the methods use \emph{exploit} strategies, which heuristically select nodes, while others utilize \emph{explore} strategies, which randomly select nodes.   High degree probing relies on the intuition that high-degree nodes in $\hat{G}$ are connected to many nodes outside of $\hat{G}$.  The structural hole methods target nodes that ``fill'' structural holes, thus connecting different parts of the graph~\cite{Burt2002Structural}.  On a similar intuition, the \textsc{CrossComm} algorithm selects nodes on the border of two communities.  Random probing selects  nodes at random from within $\hat{G}$ for probing.
 
\begin{table*}
\renewcommand{\arraystretch}{1.2}
\begin{center}
\scalebox{0.85}{
\begin{tabular}{ l  l  l  l }
Category 		& Sub-category 	& Name			& Description\\ \hline
Exploit		& Degree			& \textbf{HighDeg, LowDeg}	& Select the highest or lowest degree nodes.\\
			& Structural Hole	& \textbf{HighDisp, LowDisp}	& Select the highest or lowest dispersion nodes~\cite{Backstrom2013Dispersion}.\\
			&				& \textbf{CrossComm}		& Pick nodes with the highest fraction of neighbors outside of  \\
			& & & their community (as identified with the Louvain method~\cite{Blondel2008Louvain}). \\
			& Clustering		& \textbf{HighCC, LowCC}	& Select the highest or lowest clustering coefficient nodes.\\ \hline
Explore		&				& \textbf{Random}			& Randomly select nodes from the sample.\\

\end{tabular}}
\end{center}
\caption{\textnormal{Other popular probing methods.  We categorize methods as explore or exploit, and further subdivide as degree based, structural hole based, or random. Dispersion is an edge-based measure of how well a node's neighbors are connected to each other~\cite{Backstrom2013Dispersion}. For each node, we average the dispersion of each of its adjacent edges.}}
\label{table:strats}
\end{table*}

\noindent\textbf{Experimental Setup.}
\label{sec:setup}
For each network listed in Table~\ref{table:datasets}, we generate 20 incomplete networks using each of the four sampling methods described above.   Each incomplete network contains 10\% of the edges from the original network.  For each probing method, we conduct probes at budgets $b \in \left\{1\%, 2\%, 3\%, 4\%, 5\%\right\}$ of the number of nodes in $G$.   After conducting probes on an incomplete network $\hat{G}$, we obtain an augmented sample graph $\hat{G'}$.  To evaluate the quality of $\hat{G'}$, we simply count how many nodes it has.\footnote{Since each test graph $\hat{G}$  starts with the same number of nodes, counting the number of nodes in $\hat{G'}$ tells us how many new nodes were observed.}

\begin{figure}
\begin{center}
\begin{tabular}{cc}
\includegraphics[scale=0.29, trim=10mm 0mm 0mm 0mm]{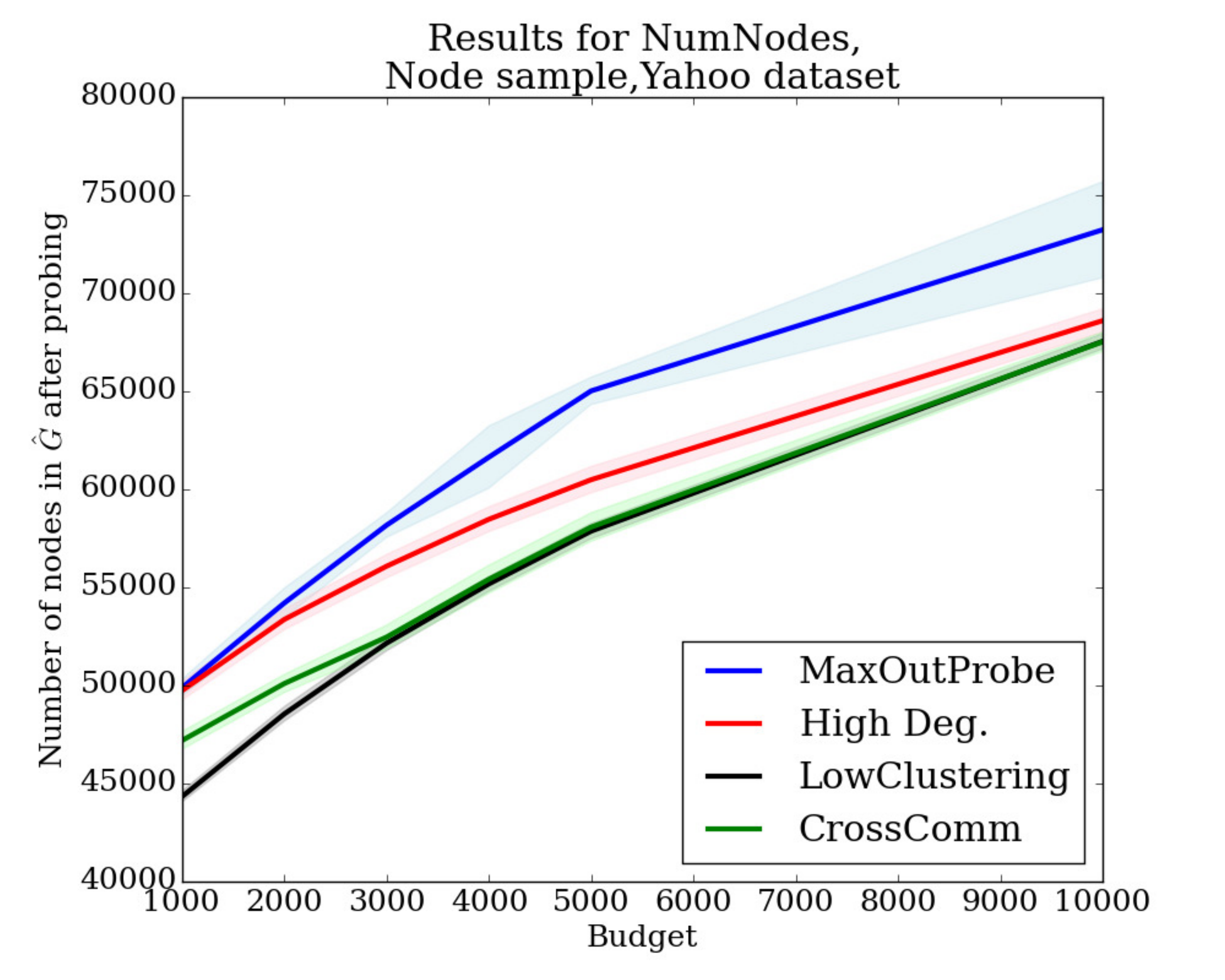} & \includegraphics[scale=0.29, trim=15mm 0mm 0mm 0mm]{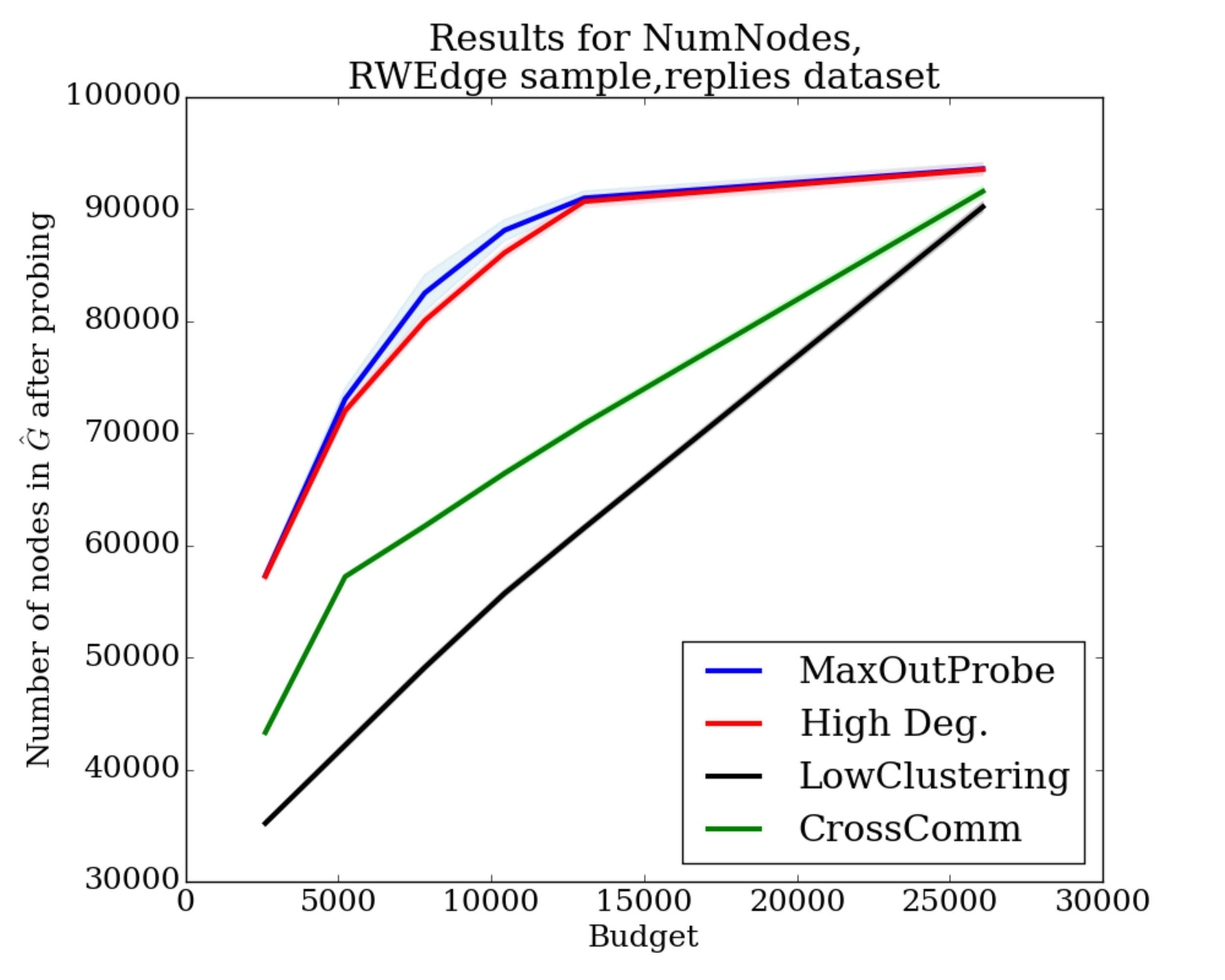} \\
(a) & (b)\\
\end{tabular}
\end{center}
\caption{Results of {\sc MaxOutProbe} vs.~popular probing methods on incomplete networks. In (a), twenty networks were observed by random node sampling on the Yahoo!~IM network.  Shading indicates one standard deviation.  Observe that {\sc MaxOutProbe} is the best across all considered node budgets. In (b), twenty networks were observed by random walk sampling on the Twitter Replies network.  Shading indicates one standard deviation.  Observe that {\sc MaxOutProbe} and High Degree probing perform similarly because the Twitter Replies network has a very low clustering coefficient (of 0.002). Results on the other datasets and sampling methods are similar to these and have been removed for brevity.}
\label{fig:YahooNodeAndRepliesRW}
\end{figure}

\noindent\textbf{Results.}
 First, we evaluate {\sc MaxOutProbe} and the other popular probing methods on the incomplete networks described above.  We observe that averaged over the 20 incomplete networks, for every sampling method and every network, {\sc MaxOutProbe} matches or outperforms the best popular probing method.  For example, see Figure~\ref{fig:YahooNodeAndRepliesRW}(a), which shows the  performances of various probing methods on incomplete networks observed by random node sampling on the Yahoo!~IM network.  In some cases, {\sc MaxOutProbe} performs roughly the same as High Degree probing (see, e.g., Figure~\ref{fig:YahooNodeAndRepliesRW}(b), for results on incomplete samples observed by random walk sampling on the Twitter Replies network).  This occurs when the incomplete network has a very low estimated clustering coefficient.   On these incomplete networks, {\sc MaxOutProbe} estimated average clustering coefficients ranges from 0.0 to 0.03.  With such a low clustering coefficient, {\sc MaxOutProbe} reduces to effectively picking the nodes with the highest observed degrees.

\begin{figure}[t]
\begin{center}
\begin{tabular}{cc}
\includegraphics[scale=0.32]{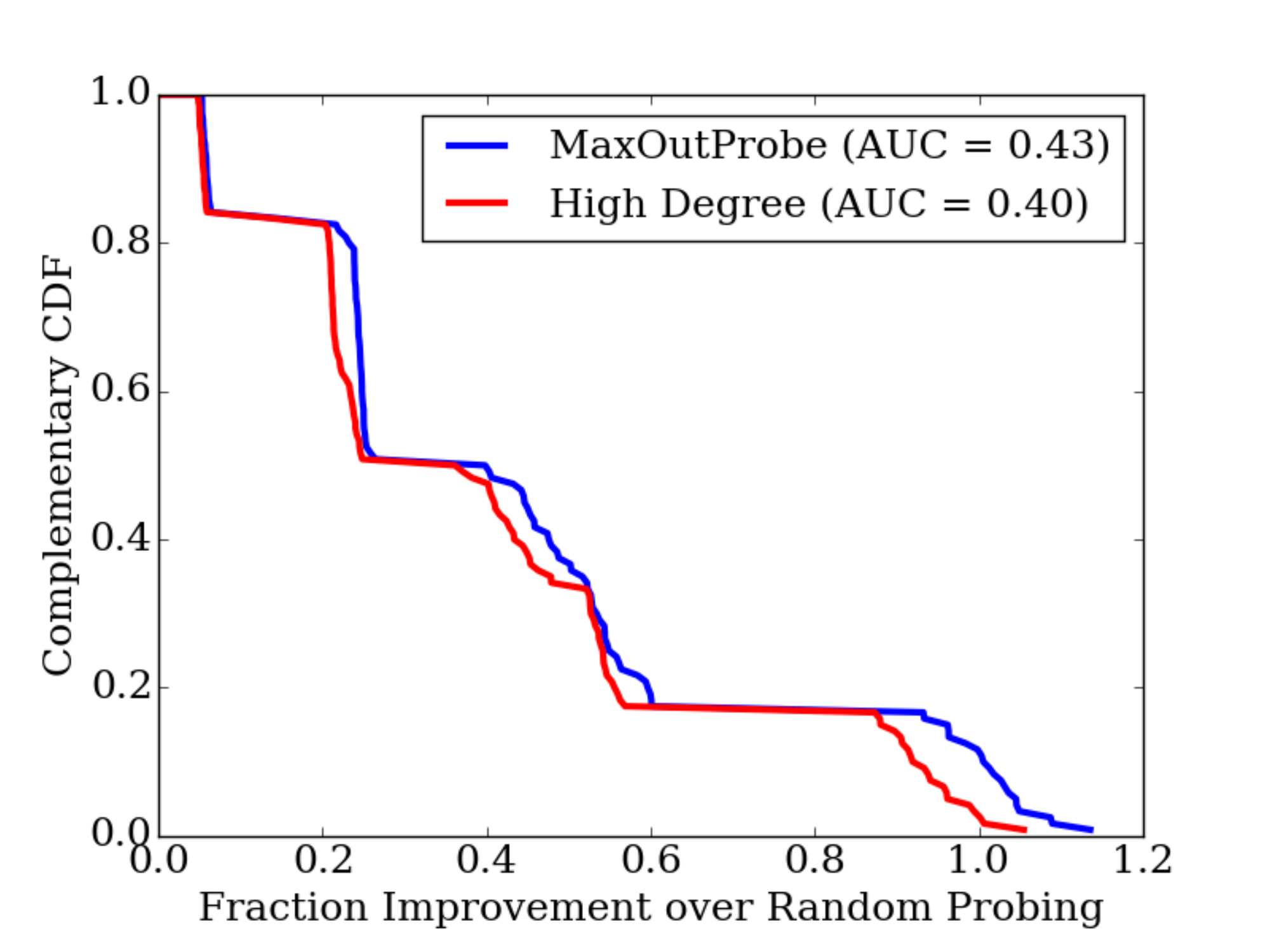}& \includegraphics[scale=0.32]{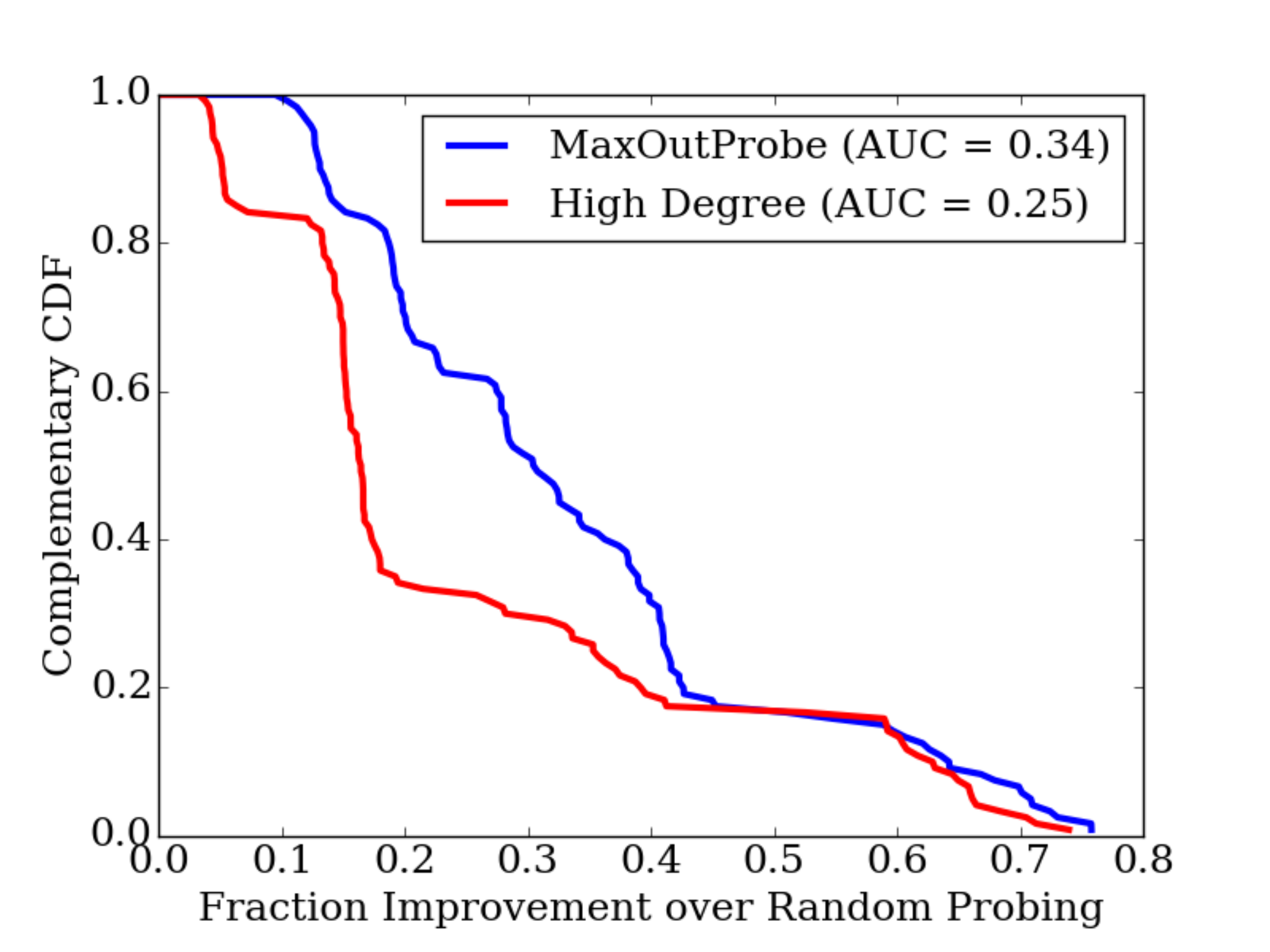} \\
(a) Observed by Random Edge Sampling & (b) Observed by Random Node Sampling\\
\includegraphics[scale=0.32]{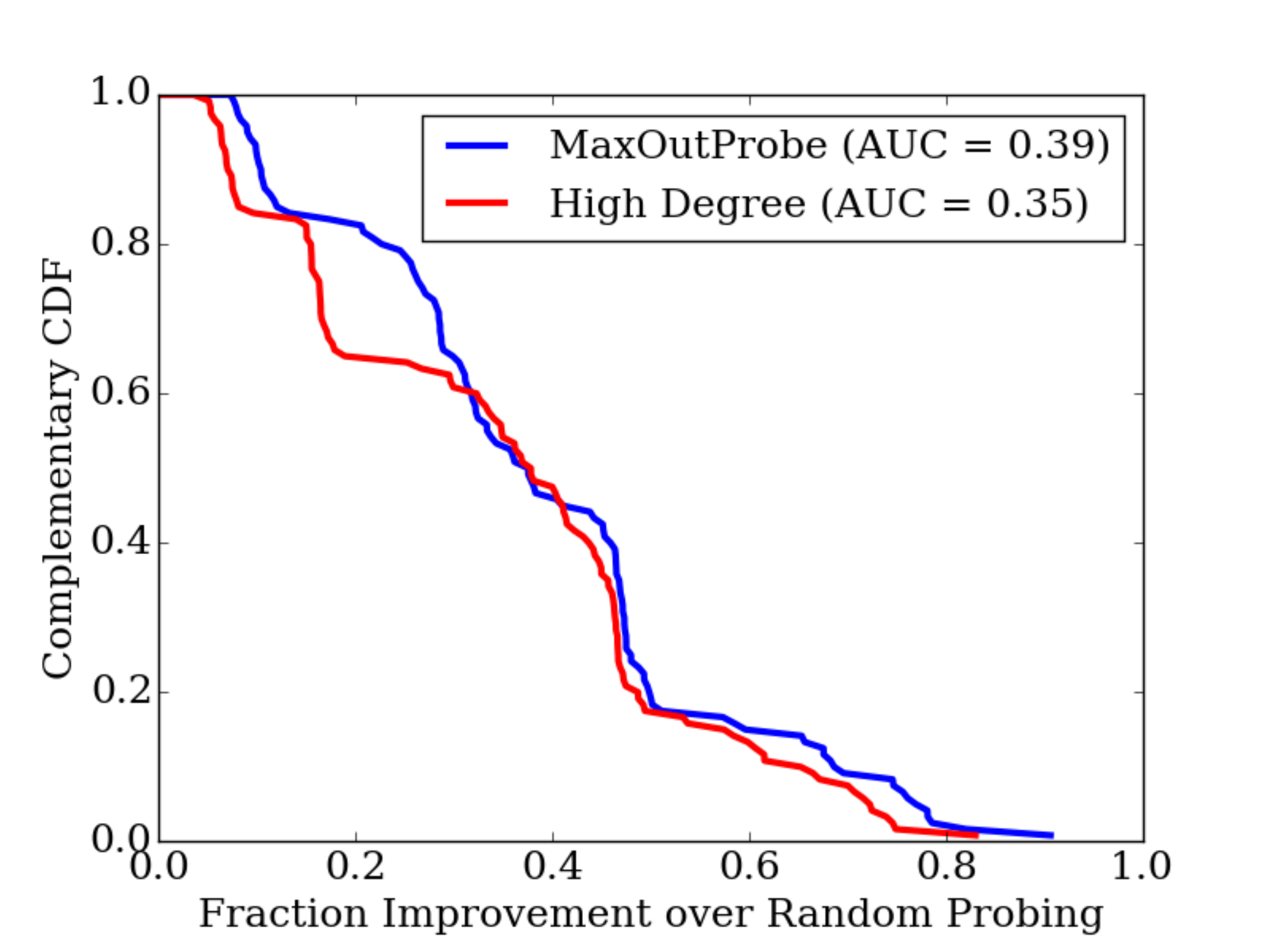} & \includegraphics[scale=0.32]{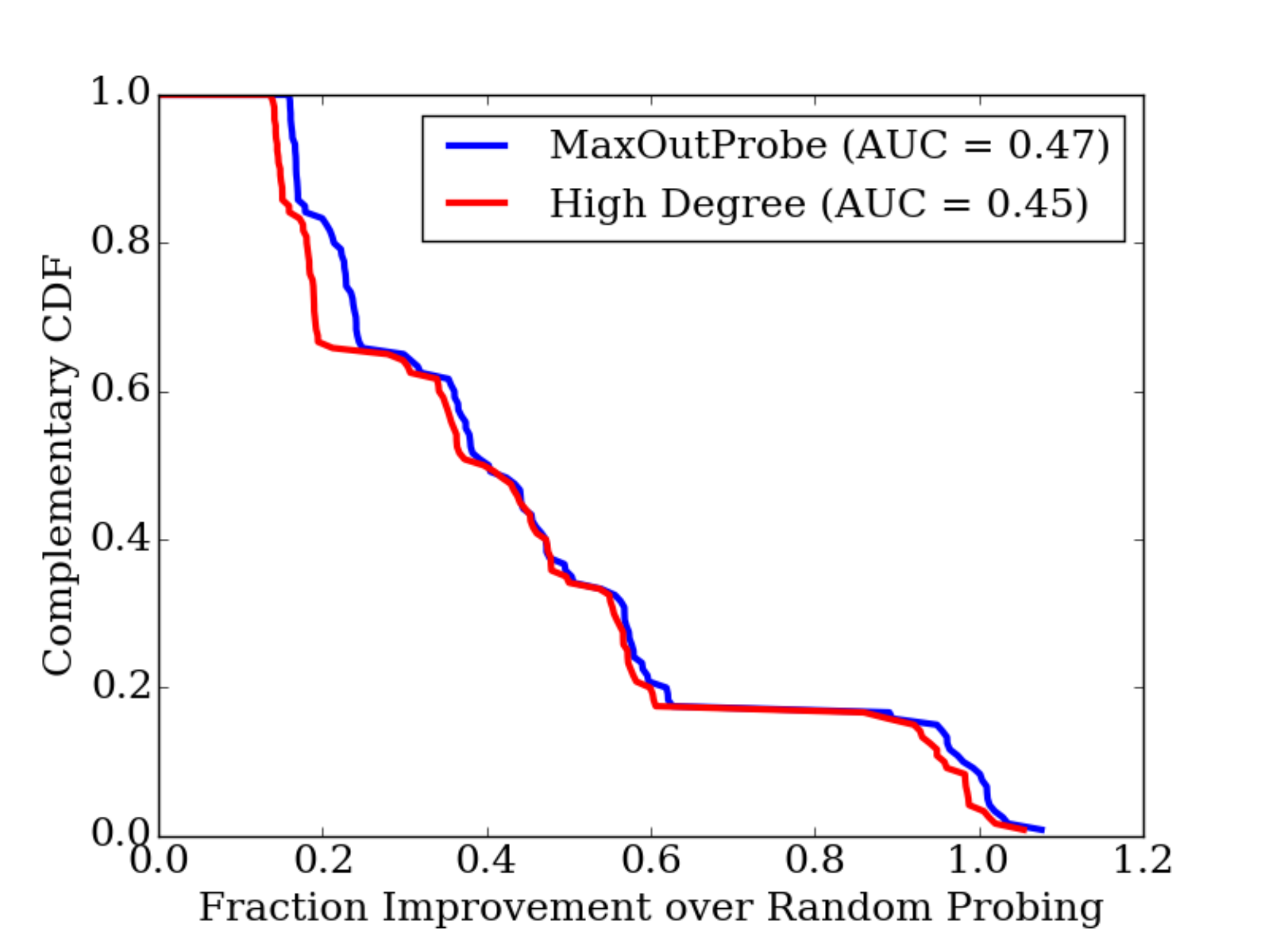} \\
(c) Observed by Random Walks & (d) Observed by Random Walks with Jumps \\
\end{tabular}
\end{center}
\caption{Complementary CDFs showing results of {\sc MaxOutProbe} and High Degree probing as compared to Random probing across all datasets, on incomplete networks observed by various sampling methods. In all cases, {\sc MaxOutProbe} has a higher area under the curve (AUC), indicating a greater improvement over Random probing than High Degree probing.}
\label{fig:AggregateResults}
\end{figure}

Rather than present plots for each network and sampling method, we aggregate results across datasets, comparing {\sc MaxOutProbe} to High Degree probing (which is typically the best popular probing method). For each incomplete network and each considered budget, we conduct probes using {\sc MaxOutProbe} and High Degree probing, and count the number of nodes in each resulting $\hat{G'}$.  Because we cannot meaningfully compare these numbers across datasets or sampling methods, we also conduct the same number of probes using Random probing, and then calculate the percent by which {\sc MaxOutProbe} and High Degree probing improved over Random probing (i.e., how much larger is the $\hat{G'}$ produced by {\sc MaxOutProbe} or High Degree probing versus that produced by Random probing?).  By comparing to Random probing, we are able to make sensible comparisons across datasets. Figure~\ref{fig:AggregateResults} shows complementary CDFs\footnote{CDF is short for cumulative distribution function.} of these aggregate results at the 0.05 probing budget on incomplete networks observed, respectively, by Random Edge sampling, Random Node sampling, Random Walk sampling, and Random Walk with Jump sampling.  Similar results were observed at other probing budgets.  The x-axis represents the percent improvement over Random probing, with values greater than 0 indicating that the approach performed better than Random probing.  The y-axis represents the fraction of such results that produced at least a certain fraction improvement over random.  We see that in all cases, {\sc MaxOutProbe} has a higher area under the curve than High Degree probing, indicating greater improvement.  The area under the curve for MaxOutProbe ranges from 4\% to 36\% higher than that of High Degree probing.

\section{Related Work}
\label{sec:related}
Our work is most related to graph sampling and crawling.  There is a rich literature on sampling graphs. For instance, previous literature has examined the study of community detection from graph samples$-$e.g., by using minimum spanning trees (MSTs)~\cite{Wu2013MST}, or by expanding a graph sample~\cite{Maiya2010Communities}.  
Avrachenkov et al.~\cite{Avrachenkov2014Degree} use queries to locate high-degree nodes.  O'Brien and Sullivan~\cite{OBrien2014Core} use local information to estimate the core number of a node. Hanneke and Xing~\cite{Hanneke2009Sample} and  Kim and Leskovec~\cite{Kim2011Sample} infer characteristics of a larger graph given a sample.   Maiya and Berger-Wolf~\cite{Maiya2013Crawling} study the problem of online sampling for centrality measures.    Cho et al.~\cite{Cho1998Crawling} study the problem of determining which URLs to examine in a Web crawl.  In contrast, we study the problem of selecting which nodes from an existing incomplete network one should probe, rather than constructing a sample from scratch.  Also, unlike many sampling methods, we are not down-sampling a network to which we have complete access, but using a limited number of probes.  

{\sc MaxOutProbe} selects probes in a batch, not incremental, manner.  This difference is critical when obtaining data is time-consuming. The closest method to {\sc MaxOutProbe} is the MEUD (Maximum Expected Uncovered Degree) algorithm by Avrachenkov et al.~\cite{Avrachenkov2014MEUD}: a sampling method intended to bring as many nodes as possible into the sample.  MEUD begins with one node; then in each step, it grows the sample by adding the node with the highest expected degree in the underlying, unobserved network.  MEUD requires knowing the degree distribution of the underlying network, but for certain classes of graphs, reduces to selecting the nodes with the highest observed degrees.  This is a valid approximation only for limited classes of random graphs, while our proposed method does not have such constraints.  Note that if we modified the MEUD algorithm to fit our problem setting, it would be reduce to High Degree probing.  
 
Researchers interested in specific network features have studied related problems.  Macskassy and Provost~\cite{Macskassy2005Guilt} address the problem of identifying malicious entities in incomplete networks by gathering more information about suspicious entities.  Cohen et al.~\cite{Cohen2003Immunization} propose an approach for immunizing a population in an unobserved network by targeting neighbors of randomly selected nodes.  Shakkottai~\cite{Shakkottai2004Sensor} considers the problem of nodes in a sensor network attempting to find the source of information. Ragoler et al.~\cite{Ragoler2004Sensor} study the problem of determining the frequency at which sensor nodes should query their environments. Lastly, graph theoreticians have studied the testability of certain graph properties (such as monotone graph properties~\cite{Alon2008Property}).

\section{Conclusions}
\label{sec:concl}
We discussed the problem of determining which nodes in an incomplete network should be probed in order to maximize the number of new nodes observed.  We presented the \textsc{MaxOutProbe} algorithm, which estimates the degree of each observed but unexplored node, as well as the graph's average clustering coefficient, to rank nodes for probing.  We demonstrated that {\sc MaxOutProbe} outperforms several popular probing methods on incomplete networks observed by four popular sampling methods over six network datasets.  Where the graph has a very low clustering coefficient, {\sc MaxOutProbe}'s performance is similar to selecting nodes with the highest observed degree.

\section*{Acknowledgments}
Soundarajan's and Eliassi-Rad's efforts were supported by Lawrence Livermore National Laboratory, by NSF CNS-1314603, by DTRA HDTRA1-10-1-0120, and by DAPRA under SMISC Program Agreement No.~W911NF-12-C-0028.  Gallagher's work was performed under the auspices of the U.S.~Department of Energy by Lawrence Livermore National Laboratory under Contract DE-AC52-07NA27344. Pinar's work was supported by the U.S.~Department of Energy, Office of Science, Office of Advanced Scientific Computing Research, Complex Interconnected Distributed Systems (CIDS) program and the DARPA GRAPHS Program, and the Laboratory Directed Research and Development program of Sandia National Laboratories. Sandia National Laboratories is a multi-program laboratory managed and operated by Sandia Corporation, a wholly owned subsidiary of Lockheed Martin Corporation, for the U.S.~Department of Energy's National Nuclear Security Administration under contract DE-AC04-94AL85000.


\section*{Appendix}
\label{sec:app}

Suppose we know the fraction of nodes $f_N$ and the fraction of edges $f_E$  of $G$ that are in $\hat{G}$;\footnote{$f_N =  \frac{\textrm{\# of nodes in } \hat{G}}{\textrm{\# of nodes in } G}$ and $f_E = \frac{\textrm{\# of edges in } \hat{G}}{\textrm{\# of edges in } G}$.} and we know that the incomplete network was observed via random node or random edge sampling.  Under such conditions, {\sc MaxOutProbe} produces unbiased estimates for the node degrees and the global clustering coefficient of the network \emph{without} having to use some of the probing budget to produce these estimates.\footnote{When {\sc MaxOutProbe} does not know how the incomplete network was observed, it estimates the average clustering coefficient of $G$ (as described earlier).  However, when {\sc MaxOutProbe} knows that the incomplete network was observed via random node or edge sampling, it estimates the total number of triangles and wedges in $G$; and uses them to produce an estimate for $G$'s global clustering coefficient.}  This section contains the proofs for these claims.

\subsection*{Estimations for Known Random Node Samples}
In random node sampling, $f_N$ fraction of the nodes in $G$ are selected at random and all their neighbors are collected.  Suppose node $u$ is in the observed network, but was not one of the nodes selected at random (that is, $u$ was observed because one of its neighbors was selected).  If node $u$ has observed degree $d^{known}_u$ in $\hat{G}$, {\sc MaxOutProbe} estimates its true degree as $\frac{1}{f_N} d^{known}_u$.  For example, if $u$ is adjacent to 5 nodes in $\hat{G}$, and 10\% of the nodes from $G$ were selected during sampling, then {\sc MaxOutProbe} estimates $u$'s true degree $d_u$ as 50.  This estimation works well for high degree nodes, but is less accurate for low degree nodes.  However, as we saw in Section~\ref{sec:exp}, probing low degree nodes in $\hat{G}$ is unlikely to lead to a large number of new nodes being observed.

\textbf{Claim \#1:} {\sc MaxOutProbe}'s estimator for $d_u$ is unbiased under random node sampling.

\textbf{Proof:}  We know that $\hat{G}$ consists of $f_N$ fraction of the nodes in $G$ selected uniformly at random, plus the neighbors of these selected nodes. Let $\hat{d_u}$ represent the estimator for $d_u$.  {\sc MaxOutProbe} sets $\hat{d_u}$ to be $\frac{1}{f_N} d^{known}_u$, where $d^{known}_u$ is the observed degree of node $u$ in $\hat{G}$. We must show that for each $u$, $\mathbb{E}[\hat{d_u}] = d_u$.

Consider a node $u$ that has not been explored during the sampling process, but is present in the sample $\hat{G}$ (that is, $u$ is adjacent to a node that was explored during sampling).  Node $u$ has $d_u$ neighbors in $G$, and because $f_N$ of the nodes from $G$ were sampled uniformly at random to produce $\hat{G}$, $\mathbb{E}[d^{known}_u] = f_Nd_u$.  Thus, $\frac{1}{f_N}\mathbb{E}[d^{known}_u] = d_u$, or equivalently $\mathbb{E}[\frac{1}{f_N}d^{known}_u] = d_u$.  Since $\hat{d_u} = \frac{1}{f_N}d^{known}_u$, we have $\mathbb{E}[\hat{d_u}] = d_u$, indicating that $\hat{d_u}$ is an unbiased estimator for $d_u$.  $\blacksquare$

To estimate the global clustering coefficient $C$ of $G$, {\sc MaxOutProbe} estimates the number of wedges (i.e., length-2 paths) and triangles in $G$.  Let $T_{\hat{G}}$ and $W_{\hat{G}}$ represent, respectively, the number of triangles and wedges in $\hat{G}$, and let $C_{\hat{G}} = 3\frac{T_{\hat{G}}}{W_{\hat{G}}}$ be the observed global clustering coefficient of $\hat{G}$.

Suppose a triangle consisting of nodes $x$, $y$, and $z$ exists in $G$.  What is the probability that it is in $\hat{G}$?  If all edges $(x, y), (y, z)$, and $(x, z)$ are in $\hat{G}$, then at least one node from each of these three edges must be fully explored during sampling.  Thus, the triangle will be in $\hat{G}$ if and only if at least two of the three nodes are selected.  We can write the probability $p_T$ that this occurs as:
\begin{equation}
\label{eq:in_neighbors}
p_T = 3f^2_N(1 - f_N) + f^3_N
\end{equation}
In other words, $p_T$ is the probability that exactly two of the three nodes are selected, plus the probability that all three are selected.  Thus, by multiplying the observed number of triangles in $\hat{G}$ by $\frac{1}{p_T}$, {\sc MaxOutProbe} estimates the number of triangles $T$ in $G$ as:
\begin{equation}
\label{eq:num_tri}
T = \frac{T_{\hat{G}}}{p_T}.
\end{equation}

We also need to calculate the probability that a length-2 path $\{(x,y), (y,z)\}$ in $G$ also appears in $\hat{G}$ ($x$ may or may not be connected to $z$).  If this path is in $\hat{G}$, either at least two nodes from $\{x, y, z\}$ were sampled, or $y$ must be sampled.  The probability $p_W$ of this occurring is as follows:
\begin{equation}
\label{eq:in_neighbors}
p_W = f^3_N + 3(f^2_N(1 - f_N)) + f_N(1-f_N)^2
\end{equation}
This is the probability that all three nodes are sampled plus the probability that exactly two nodes are sampled plus the probability that just $y$ is sampled.  {\sc MaxOutProbe} then multiplies the observed number of length-2 paths from $\hat{G}$ by $\frac{1}{p_W}$ to estimate the number of wedges $W$ in $G$ as:
\begin{equation}
\label{eq:num_wedges}
W = \frac{W_{\hat{G}}}{p_W}.
\end{equation}

$G$'s global clustering coefficient  is estimated as $\hat{C}=3\frac{T}{W} = 3(\frac{T_{\hat{G}}}{p_T})/(\frac{W_{\hat{G}}}{p_W}) = 3\frac{p_W}{p_T}\frac{T_{\hat{G}}}{W_{\hat{G}}} = \frac{p_W}{p_T}C_{\hat{G}}$.

\textbf{Claim \#2:} {\sc MaxOutProbe}'s estimator for $C$ is an unbiased estimate of the graph's true global clustering coefficient under random node sampling.  

\textbf{Proof:} $\hat{G}$ consists of $f_N$ fraction of the nodes in $G$ selected uniformly at random, plus the neighbors of these selected nodes. Let $\hat{C}$ represent the estimator for $G$'s true global clustering coefficient $C$.  {\sc MaxOutProbe} sets $\hat{C}$ to be $\frac{p_W}{p_T}C_{\hat{G}}$, as described above. We must show that $\mathbb{E}[\hat{C}] = C$.  

Suppose that the wedge $\{(x, y), (y, z)\}$ appears in $\hat{G}$.  Also, suppose that the nodes $x$, $y$, and $z$ form a triangle in $G$ (i.e., edge $(x, z)$ is present in $G$).  What is the probability that edge $(x, z)$ is present in $\hat{G}$ (i.e., the nodes $x$, $y$, and $z$ form a triangle in $\hat{G}$)?  Because we assume that the wedge $\{(x, y), (y, z\}$ is present in $\hat{G}$, some of these three nodes must have been selected during the sampling process.  Thus, we need to consider three  possibilities: 

\begin{enumerate}

\item All three of $x, y, z$ were sampled. The probability of this occurring is $f_N^3$.

\item Exactly two of $x$, $y$, and $z$ were sampled.  The probability of this occurring is $3f_N^2(1 - f_N)$.

\item Only $y$ was sampled.\footnote{Recall that we observe the wedge $\{(x, y), (y, z)\}$ in $\hat{G}$.}  The probability of this occurring is $f_N(1 - f_N)^2$.   
\end{enumerate}

In possibilities (1) and (2), edge $(x, z)$ will certainly be present in $\hat{G}$.  In possibility (3), edge $(x, z)$ will not be present in $\hat{G}$.  Thus, given that the wedge $\{(x, y), (y, z)\}$ is present in $\hat{G}$, the probability $P_{closed}$ that edge $(x, z)$ is also present in $\hat{G}$ is as follows:
\begin{equation}
P_{closed} = \frac{f_N^3 + 3f_N^2(1 - f_N)}{f_N^3 + 3f_N^2(1 - f_N) + f_N(1 - f_N)^2} = \frac{p_T}{p_W}
\end{equation}  

Thus, given that a set of nodes forms a wedge in $\hat{G}$ and a triangle in $G$, $P_{closed}$ is the probability that it also forms a triangle in $\hat{G}$.  In other words, $\mathbb{E}[C_{\hat{G}}] = P_{closed}{C} = \frac{p_T}{p_W}{C}$.  Since the definition of $\hat{C}$ is simply $\frac{p_W}{p_T}C_{\hat{G}}$, we have that $\mathbb{E}[\hat{C}] = C$, so $\hat{C}$ is an unbiased estimator of $C$. $\blacksquare$

\subsection*{Estimations for Known Random Edge Samples}

In random edge sampling, $f_E$ fraction of edges are sampled uniformly at random.  To estimate a node's true degree $d_u$, {\sc MaxOutProbe} multiplies its observed degree $d^{known}_u$ by $\frac{1}{f_E}$.  As with random node sampling, this estimation is inaccurate for low degree nodes, but such nodes are unlikely to be selected, and this inaccuracy does not affect {\sc MaxOutProbe}'s selection of nodes to probe.  

\textbf{Claim \#3:} {\sc MaxOutProbe}'s estimator for $d_u$ is unbiased under random edge sampling.  

\textbf{Proof:} We know that $\hat{G}$ consists of a fraction $f_E$ of the edges in $G$ selected uniformly at random.  Let $\hat{d_u}$ represent the estimator for $d_u$.  {\sc MaxOutProbe} sets $\hat{d_u}$ to be  $\frac{1}{f_E}d^{known}_u$, where $d^{known}_u$  is the observed degree of node $u$ in $\hat{G}$. We must show that for each $u$, $\mathbb{E}[\hat{d_u}] = d_u$. 

Consider a node $u$ in $\hat{G}$, with true degree $d_u$.  Because $f_E$ of the edges from $G$ were sampled uniformly at random, $\mathbb{E}[d^{known}_u] = f_Ed_u$, or equivalently $\mathbb{E}[\frac{1}{f_E}d^{known}_u] = d_u$.  Therefore, $\mathbb{E}[\hat{d_u}] = d_u$, indicating that $\hat{d_u}$ is an unbiased estimator for $d_u$. $\blacksquare$

Consider a length-2 path $\{(x, y), (y, z)\}$ in $G$ that is present in $\hat{G}$.  If $x$ is connected to $z$ in $G$ (i.e., there exists a triangle between $x$, $y$, and $z$ in $G$), then with probability $f_E$, edge $(x, z)$ is also present in $\hat{G}$.  Thus, {\sc MaxOutProbe} estimates $C$ (the global clustering coefficient of $G$) to be $\frac{1}{f_E}C_{\hat{G}}$, where $C_{\hat{G}}$ is the global clustering coefficient of the incomplete network $\hat{G}$. 

\textbf{Claim \#4:} {\sc MaxOutProbe}'s estimator for $C$ is unbiased under random edge sampling.

\textbf{Proof:} We know that $\hat{G}$ consists of a fraction $f_E$ of the edges in $G$ selected uniformly at random. Let $\hat{C}$ represent the estimator for $C$.  {\sc MaxOutProbe} sets $\hat{C}$ to be $\frac{1}{f_E}C_{\hat{G}}$, where $C_{\hat{G}}$ is the global clustering coefficient of the incomplete network $\hat{G}$. We must show that $\mathbb{E}[\hat{C}] = C$.  

$C$ is the fraction of wedges (i.e., length-2 paths $\{(x, y), (y,z)\}$) that are closed (i.e., $x$ is connected to $z$).  Suppose that the edge $(x, z)$ exists in $G$. Then, independently of the wedge $\{(x, y), (y,z)\}$ being present in $\hat{G}$, there is $f_E$ probability that $(x, z)$ is present in $\hat{G}$.  Thus, of the wedges in $G$ that are present in $\hat{G}$, we expect that $f_EC$ fraction of these wedges are closed in $\hat{G}$. If $C_{\hat{G}}$ is the global clustering coefficient of $\hat{G}$, then $\mathbb{E}[C_{\hat{G}}] = f_EC$, or equivalently $\mathbb{E}[\frac{1}{f_E}C_{\hat{G}}]= C$.  Therefore, $\mathbb{E}[\hat{C}] = C$, indicating that $\hat{C}$ is an unbiased estimator for $C$.  $\blacksquare$

\subsection*{Remarks} The aforementioned proofs (for both random node and random edge sampling) show that the estimates are correct in expectation, but do not provide concentration bounds. By applying Hoeffding's inequality~\cite{hoeffding1963}, we can bound the errors in expectations. While these bounds may be weak for each node, they  provide a theoretical basis for showing that we can identify all high-degree nodes up to some approximation (see Theorem 6.2 in~\cite{ThBe13}).

\end{document}